\documentclass[conference,compsoc]{IEEEtran}
\usepackage[dvips]{graphicx}
\usepackage{cite}
\usepackage{color}
\usepackage[cmex10]{amsmath}
\usepackage{amssymb}
\usepackage{amsmath}
\usepackage{booktabs}
\usepackage{multirow}
\usepackage{dcolumn}
\usepackage{amsthm}
\usepackage{enumerate}
\usepackage{algorithmicx}
\usepackage{array}
\usepackage{mdwmath}
\usepackage{mdwtab}
\usepackage{eqparbox}
\usepackage{verbatim}
\usepackage{bm}

\usepackage[utf8]{inputenc}
\usepackage{url}
\usepackage{enumitem}
\usepackage{xcolor}
\usepackage[]{footmisc}

\usepackage{algorithm}
\usepackage{algpseudocode}
\usepackage{color,soul}

\newtheorem{remark}{Remark}
\newtheorem*{remark*}{Remark}

\IEEEoverridecommandlockouts

\def \tclb{\textcolor{black}}
\def \tclm{\textcolor{black}}
\def \tclo{\textcolor{black}}
\def \tclg{\textcolor{black}}

\graphicspath{{figs/}}

\makeatletter
\newcommand*\titleheader[1]{\gdef\@titleheader{#1}}
\AtBeginDocument{%
    \let\st@red@title\@title
    \def\@title{%
        \vspace*{-3.0em}%
        \bgroup\normalfont\large\centering\@titleheader\par\egroup
        \vskip2.0em\st@red@title}
}
\makeatother
\titleheader{\fontsize{9}{\baselineskip}\selectfont{Best Paper Award Recipient at the 6th Silicon Valley Cybersecurity Conference (SVCC 2025)}}
\title{Helol Tunnel: Covert Channel Exploitation of TLS Extensibility \& Privacy Features}

\begin{document}


 \author{
    \IEEEauthorblockN{
        Reza Soosahabi,\thanks{Corresponding author: reza.soosahabi@keysight.com. Project artifacts are available at: \texttt{https://github.com/Keysight/helol-Tunnel}}
        Rakesh Seal
    }
    \IEEEauthorblockA{
        Application \& Threat Intelligence Research Center\\
        Keysight Technologies, Inc.\\
        Austin, TX, USA
    }
}

\maketitle

\begin{abstract}

Covert channels exploiting network protocols for data exfiltration and command-and-control (C2) are integral parts of modern cyberattacks.
In search of a significant covert channel within the fabric of the Internet, we targeted the combinatorial properties of the Client Hello (CHLO) packets in the ubiquitous Transport Layer Security (TLS) protocol.
The proposed Helol tunnel is a novel covert approach to embedding information in TLS Client Hello packets, which involves the strategic rearrangement of their cryptographic information elements.
To sustain TLS protocol extensibility, the recent anti-ossification TLS compliance measures encourage the interactive middleboxes and next-generation firewalls (NGFWs) to preserve the parameter configuration in the Client Hello packets.
Furthermore, to improve user privacy, popular Internet applications are varying their TLS CHLO parameter configurations to resist TLS fingerprinting by third-party network entities.
We demonstrate the strength of the Helol tunnel to exploit these recent developments to evade NGFWs with interactive proxy and comprehensive threat protection.
We also numerically show the efficacy of Helol tunneling over state-of-the-art covert channels that exploit TLS through the use of real traffic captures and public TLS fingerprinting data.


\end{abstract}

\begin{IEEEkeywords}
covert channel, malware, data exfiltration, middleboxes, TLS fingerprinting, TLS ossification, network security
\end{IEEEkeywords}

\IEEEpeerreviewmaketitle
\section{Introduction}
Once infiltrating their victim networks, most modern malware leverage covert channels to secretly communicate (callback) with a remote attacker while bypassing the security barriers. Maintaining persistent covert channels is crucial for attackers to command and control (C2) their malware agents, which are designed to commit malicious activities such as exfiltration of sensitive data, launching distributed denial of service (DDoS) attacks, deploying ransomware and mining cryptocurrency \cite{apt2}. To protect networks against such threats, modern next-generation firewalls (NGFWs)  have evolved to detect and block these covert channels in addition to defending against malware infiltration.

\begin{figure}[t] \centering
    \includegraphics[width=9cm]{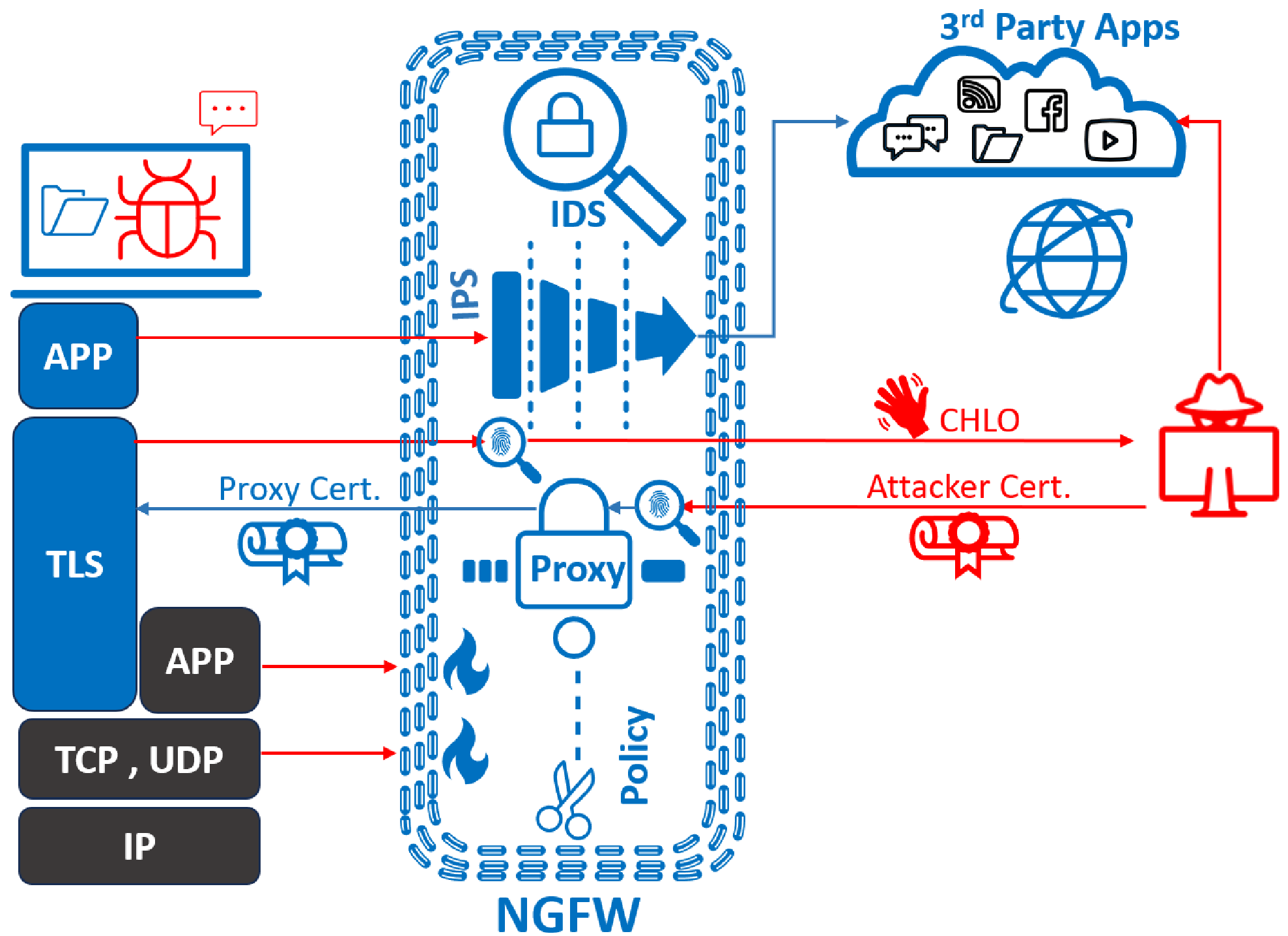}
		\caption{Next Generation Firewall (NGFW) with IPS/IDS}
    \label{ngfw-fig}
\end{figure}

There has been extensive research on covert channel techniques for exploiting network protocols at different layers. In Section \ref{ngfw-sec}, we start with the security features of modern NGFWs, including blocking covert channels that exploit layer 3 \& 4 protocols. Most internet applications, including modern malware, utilize the HTTPS application protocol, which uses TLS for server authentication and data encryption. We continue to detail the NGFW TLS decryption mechanism, as well as effective countermeasures against covert channels that exploit third-party HTTPS applications. Then we review the recently disclosed covert channel methods that encode data in the first packet of the TLS protocol handshake (i.e., the Client Hello (CHLO) message). NGFWs must relay this message to the associated destination server before initiating the data decryption and inspection process.  We dedicate Section \ref{ja3-sec} to an expanded discussion of TLS CHLO features and their importance in application fingerprinting and network forensics. We detail two recent developments in this area that motivated our research. Internet applications fingerprinting evasion by randomizing TLS CHLO parameters, and anti-ossification standard recommendations to limit the ability of firewalls and interactive middleboxes to alter TLS CHLO parameters.

In Section \ref{helol-sec}, we present our proposed covert channel method, \textit{Helol tunnel}, in its common exploitation scenario, where malware agents use it to exfiltrate sensitive data to its C2 server in the form of TLS CHLO packets barely distinguishable from the incumbent TLS traffic. As reflected in its name, our proposed method exploits the permitted combinatorial arrangement (selection \& permutation) of the cryptographic parameters in the TLS CHLO for the purposes of covert message encoding. Benefiting from recent developments in the realm of TLS compliance and fingerprinting, our proposed Helol tunneling method achieves the following advantages over the best known covert channels of its kind:
\begin{enumerate}
	\item Evading NGFWs with interactive TLS proxy complying with anti-ossification measures such as \cite{rfc8701}
        \item No reliance on third-party web applications
        \item Indistinguishable from the normal application traffic
	\item No reliance of malware on TLS software libraries
	\item Can coexist with other TLS covert channel methods
	\item Capable of passive (IP-less) channels across NGFWs \label{pcci}
	\item Throughput gains over TLS CHLO covert channels \cite{snicat}\label{big-gain}
\end{enumerate}

Potential remediation strategies against the Helol tunnel are outlined in Section \ref{remediation-sec}, followed by a presentation of throughput computations and numerical results that support the gains mentioned in item \ref{big-gain} above. The concluding remarks are presented in Section \ref{conc-sec} followed by the appendices.

Appendix \ref{app-a} is dedicated to analyzing the efficacy of common combinatorial encoding schemes used in this work.
\tclm{In Appendix \ref{fw-results}, we present the results indicating how our proposed method can evade some of the commonly deployed enterprise NGFWs in a laboratory setting.}
We have dedicated Appendix \ref{pcc-sec} to briefly expand on the passive (IP-less) exploitation scenario and its importance for future supply-chain security research.
Finally, we have addressed the ethical consideration and responsible disclosure of this research in the Appendix \ref{ethics}.

\section{Next-Generation Firewalls Defense Against Covert Channels} \label{ngfw-sec}

In the recent decade, the network security landscape has changed toward the defense-in-depth model (DiD) with an elaborate system of policies to intercept and monitor external network access at different layers. Our work involves covert channels capable of compromising the most protected network segments (slices) that host its most sensitive assets. Looking at the illustration in Figure \ref{ngfw-fig}, the NGFWs concentrate all their security features on the IP traffic traversing these segments to the Internet, including interactive proxy, intrusion detection / protection systems (IDS / IPS), application layer data loss prevention (DLP) and more \cite{cloudflare-did}.

The theory of covert communication dates back to the early research on information security, such as reference \cite{lampson1973}.
The famous prisoners' problem in \cite{subliminal-1984} shaped the common nomenclature for the covert channel research that extends to the network security: the alleged captives, Alice (malware agent) and Bob (C2 server), wish to exchange messages without the warden (firewall) detecting and tampering with them. The survey by Wendzel et al. in \cite{wendzel-survey} offers a well-structured catalog of covert channels in various protocols that reveals similar patterns of weaknesses at different network layers. Using the terminology of the OSI model, the methods using layers 1 and 2 often work in the wireless domain without the need to access the Internet \cite{reza-sparrow}
. Moving up the stack to layers 3 and 4, there has been tremendous research, since the early days of the Internet, to exploit the features of TCP, UDP, and IP protocols such as ICMP for covert channels \cite{zander-survey-2007, tcp-exfil,covert-tcp-survery, len-stego-udp}.
Present day NGFWs have default features designed to thwart covert channel exploitation in these layers. Example features include explicit proxy and carrier grade network address translation (CGNAT), which prevent direct sockets from internal hosts to outside Internet hosts.

Most of the common layer 7 protocols, such as HTTPS, are secured by Transport Layer Security (TLS) across the Internet. Most malware covert channels have been reported to use applications running over TLS \cite{malware-tls}. Hence, to decrypt and analyze application payloads by IPS / IDS modules, the NGFW proxy intercepts and interposes itself in each TLS connection, thereby creating two separate TLS connections. As shown in Figure \ref{ngfw-fig}, the client sends the TLS Client-Hello (CHLO) packet with its supported cryptographic parameters, which the NGFW inspects and relays to the respective destination server. The client authenticates the server before establishing a session encryption key. The destination server replies with its public key in a certificate format (signed by a certificate authority). The NGFW inspects the authenticity of the server certificate and uses it to establish its own TLS connection with the server, thereby masquerading as the client. It then passes its own trusted certificate to the client, masquerading as the server, to establish the connection.
\begin{figure}[t] \centering
    \includegraphics[width=9cm]{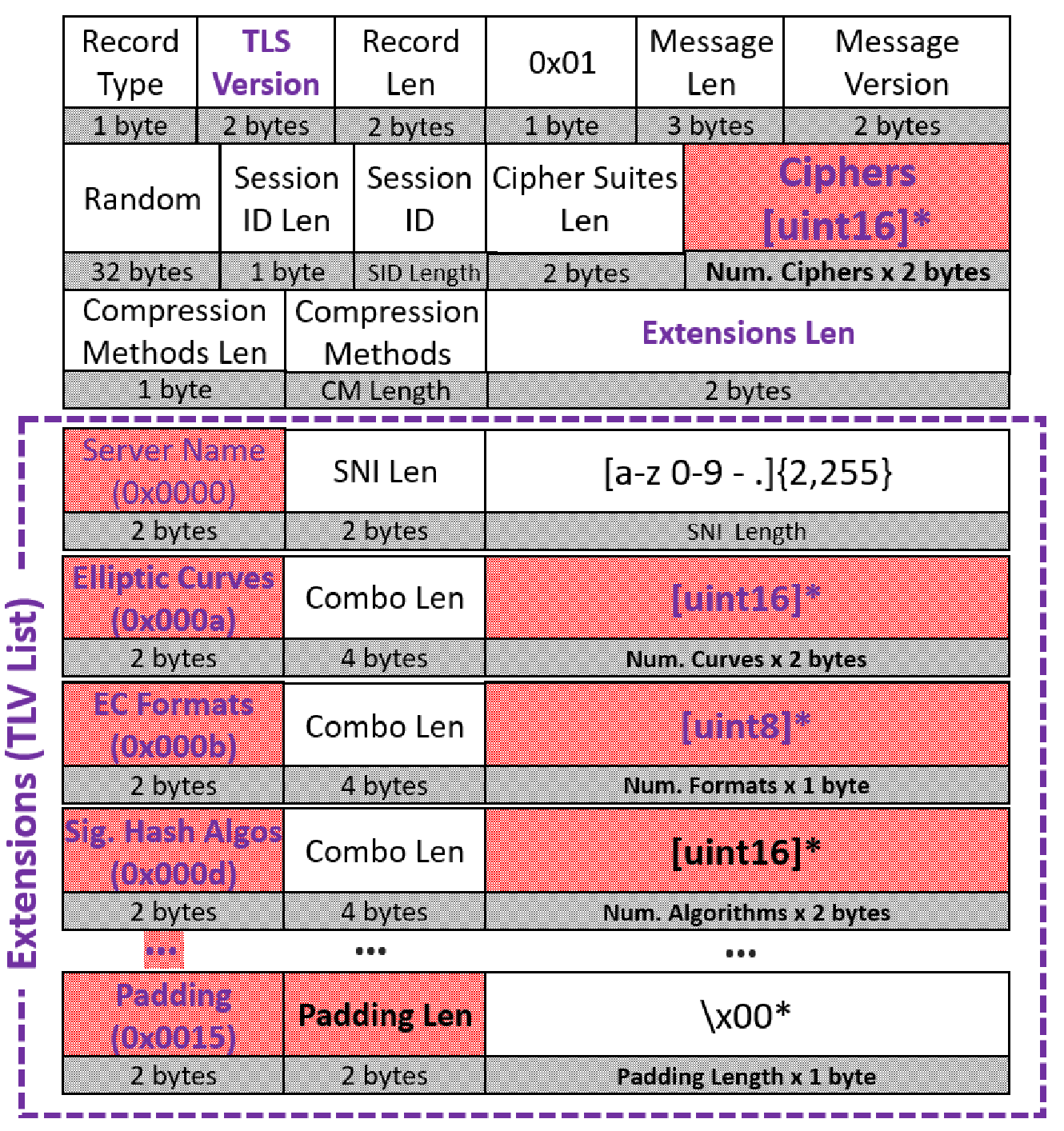}
		\caption{TLS Client Hello (CHLO) payload structure.}
    \label{chlo-fig}
\end{figure}

Foregoing the costs and challenges of obtaining authorized certificates, a large fraction of malware agents have resorted to third-party HTTPS apps, especially file-sharing apps, to establish an indirect covert channel with their C2 servers. These covert channels leverage steganographic techniques to obfuscate stolen data in files or messages uploaded to a third-party application server and later accessed by an attacker \cite{mitre2019command, mitre2020web}. However, the security industry has developed robust solutions against these channels: (1) NGFWs can now optionally restrict access to such third-party services; and (2) application / cloud providers have developed several strategies to root out C2 activities and grant access only to verified applications and users. These strategies may include human verification and continuous turnover of their API methods \cite{infosecurity2023battling}.

\tclo{
Reexamining the illustration in Figure \ref{ngfw-fig}, the content of a TLS CHLO packet from an infected host can still reach a C2 server whose authenticity has not yet been verified through the interactive proxy process. Covert channels that exploit the features of the TLS protocol are gaining traction in the security research community. Heinz et al. in \cite{heinz-tls-1, heinz-tls-2} have identified several potential TLS covert channel schemes, such as embedding data in the TLS CHLO initialization vector,  and acknowledged their limitation to overcome interactive proxies as effective countermeasures. Methods using multiple packets, such as modulating data over successive TLS messages of different record type, yield very low throughput compared to TLS CHLO methods and could lead to traffic pattern anomalies detectable by NGFWs. There are some early techniques to exploit the optional extension fields in TLS CHLO or in X.509 Client certificates for data exfiltration that had limited success against the NGFW TLS inspection \cite{stop-certexfil}. The methods exploiting the widely supported TLS extensions had better chances against NGFWs. Being a successful example, the toolkit called \textit{SNICat} targets the application-defined server name extension (SNI) in TLS CHLO for data exfiltration \cite{snicat}. Similarly to the classic DNS tunneling \cite{dns-tun-survery}, it encodes data in the alphanumeric characters of the server name. The throughput for the SNICat channel has been analyzed in Section \ref{tput} in comparison to our proposed method. It should be noted that SNICat data encoding also leads to anomalous SNI patterns that the major NGFWs are already equipped to detect and block using the same methods used to combat DNS tunneling and URL filtering \cite{pan-resp}.
}

\section{TLS Fingerprinting \& NGFW Compliance} \label{ja3-sec}
TLS CHLO packets conform to ASN.1 BER encoding that contains several type-length-value (TLV) encoded data elements starting with the cipher list. Depicted in Figure \ref{chlo-fig}, it can also have a varying number of extension fields which have their own TLV encoding. The Internet Assigned Numbers Authority (IANA) maintains the list of valid TLS CHLO parameters and extensions. It turns out that the parameters used in the TLS CHLO can be used to identify the client software stack, much like the HTTP user-agent string, but without incurring the decryption overhead.

There has been a growing interest in fingerprinting TLS CHLO parameters for network forensics, anomaly detection, malware C2 detection, and even DDoS protection against botnets \cite{ja3-springer}. Developed by SalesForce in 2017, JA3 is the most widely adopted method for fingerprinting (by MD5 hash) a set of important TLS CHLO parameters, which appear in purple font in Figure \ref{chlo-fig}, with their respective encoding format. The order of the TLS extensions can be changed at any point, and the client has full discretion over the order of its preferred cipher suites, elliptic curves (supported groups) and the other cryptographic parameters, as long as the receiving server at least finds a supported value in each list. Since the process of deriving the JA3 hash is sensitive to the ordering of its components, it has become a common trend among malware and even web browsers advocating for user privacy to randomize their JA3 by moving parameters around. As the industry accepts shuffling of the TLS CHLO parameters as a common privacy protection practice, new methods of sort-before-hash fingerprinting, such as JA4, are more stable and are gaining popularity \cite{vectra-ja3}. \tclo{There are also proposals to circumvent any fingerprinting attempts by full encryption of TLS CHLO such as in \cite{ech-draft} that are under review by NGFW providers due to their potential for abuse. \cite{ech-broadcom}.}

In general terms, \textit{protocol ossification} occurs when intermediary network middleboxes, such as NGFWs, enforce strict protocol behaviors, thereby limiting the extensibility and future updates of network protocols\cite{ieee-ossif, rfc9170}. Historically, some NGFWs have modified cryptographic parameters in TLS CHLO packets before forwarding them to their destination servers.
\tclm{To maintain control over the security of their products, the Internet application developers and the cloud service providers have been campaigning for TLS anti-ossification measures by standardization bodies such as the IETF.
Being the prominent example, the proposed GREASE (Generate Random Extensions and Sustain Extensibility) TLS feature \cite{rfc8701} was ratified in RFC 8701. It allows Internet applications to use IANA-registered unused extensions and cipher codes to detect and react to ossifying TLS middleboxes. Furthermore, some application providers use the TLS certificate pinning techniques in their client software to prevent interception by interactive TLS proxies. Enterprise NGFW operators can configure their desired list of trusted hosts for proxy exclusion. \cite{pinning}.}
\begin{remark}\label{motiv}
\tclb{The methods of Internet applications to avoid JA3 fingerprinting and the anti-ossification measures to preserve the configuration of the TLS CHLO parameters motivated us to develop a new covert channel named the “Helol Tunnel”. As reflected in its name, this method can evade NGFWs in a more subtle way, by encoding messages in the arrangement (permutation or selection) of list-type TLS CHLO parameters. It can withstand compliant interactive proxies and be more effective than the other methods encoding data in the TLS extension.}
\end{remark}

\section{Proposed TLS CHLO Covert Channel} \label{helol-sec}
Attackers can use the Helol tunnel for various types of C2 communication. Here we explain the exfiltration channel scenario as a generalizable example. Reviewing the taxonomy of covert channel methods presented in \cite{wendzel-survey}, there is a literature on the use of sequence patterns. However, most of these works fall in layer 3 \& 4 which, as discussed in Section \ref{ngfw-sec}, lose their potency against modern NGFWs. The authors in \cite{http-headers, http-perm} presented sequence-based covert channel opportunities using the HTTP protocol. The former suggests using the order of HTTP headers, while the latter involves reordering of the HTTP request responses. Assuming an upgraded HTTPS environment, these methods are in the lower-throughput category and may barely survive NGFW heavy inspection of HTTP headers and the modern HTTP/2 \& 3 dynamics.
\begin{figure}[ht] \centering
    \includegraphics[width=8.8cm]{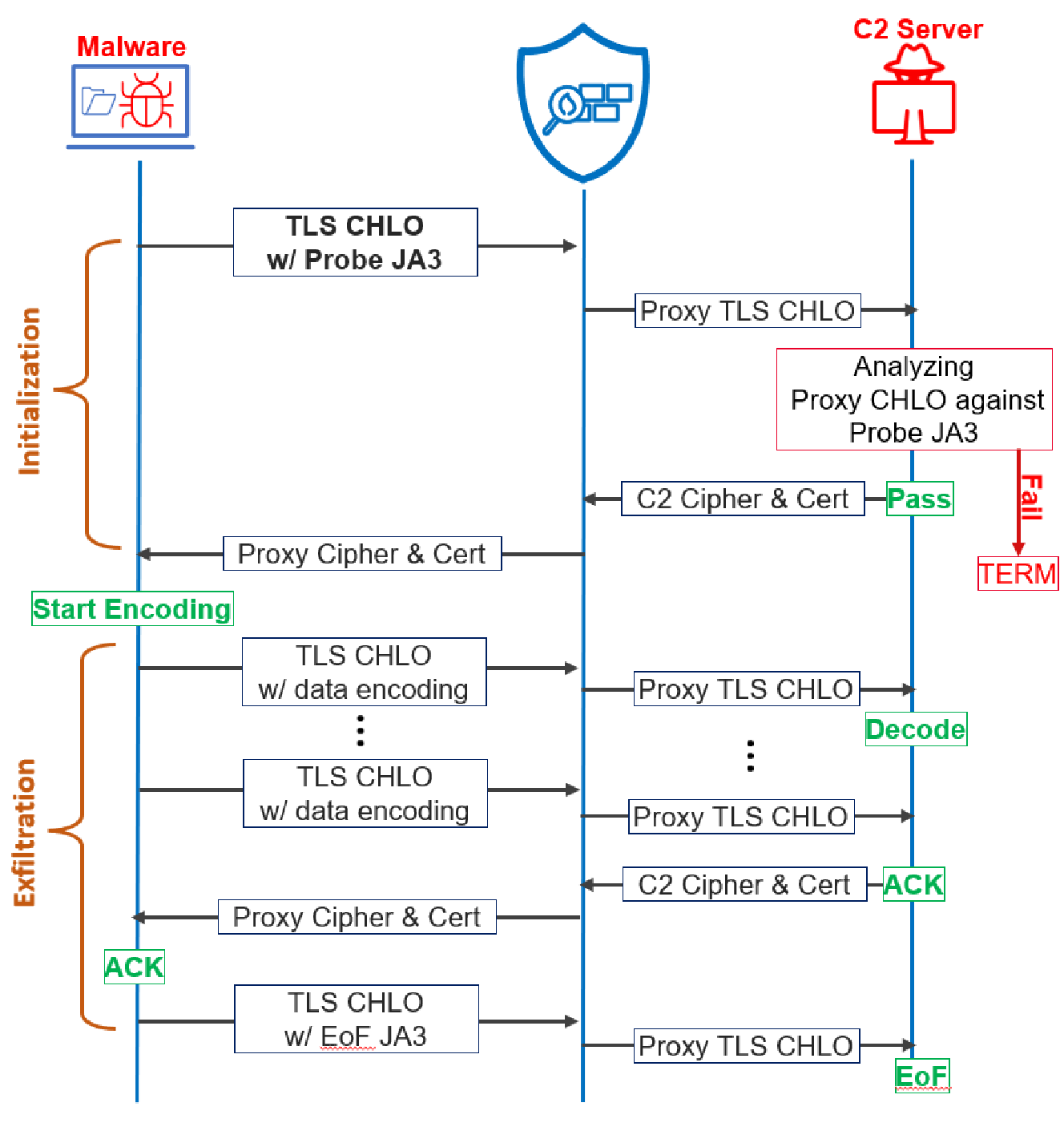}
		\caption{Common Helol Tunnel exploitation scenario}
    \label{active-diag}
\end{figure}
Solely exploiting the TLS CHLO packets, the Helol tunnel sender-receiver synchronization mechanism bears similarity to that presented in \cite{snicat}. Illustrated in Figure \ref{active-diag}, we explain how a malware-infected host in the local network establishes a synchronized covert channel with a remote C2 server through an enterprise NGFW with interactive proxy. In this data exfiltration scenario, the malware agent requires more uplink bandwidth for data exfiltration, and minuscule downlink bandwidth for receiving commands and acknowledgments (ACKs) from the C2 server.
Upon receiving the TLS CHLO through the NGFW, the C2 server can implement the ACK / NACK function in several ways, including: (1) responding with a valid certificate or closing the connection, (2) responding with valid or invalid certificate as in \cite{snicat}, (3) using the choice of cipher choice. We believe that the Helol tunnel uplink channel is compatible with other alternative downlink signaling channels that are not covered here.

\subsection{NGFW Probing \& Initialization}
\tclm{
The installed malware is shipped with a lightweight data encoding scheme to generate a TLS CHLO packet with the desired arrangement of ciphers, extensions, elliptic curves and hashing algorithms, colored red in Figure \ref{chlo-fig}.} It should be noted that there is no actual encryption that is happening in our proposed method, as TLS handshakes do not have to advance past the server response shown in Figure \ref{active-diag}. Therefore, the malware agent can generate TLS CHLO packets directly as TCP payloads without invoking SSL libraries on the victim host.

\tclm{ Delineated in Figure \ref{active-diag}, the Helol tunnel starts with a probing phase (initialization) where the malware agent sends the first TLS CHLO packet to detect ossifying middleboxes.
The values in the initial TLS CHLO are configured to produce common JA4 fingerprints from common Internet applications.
Each TLS CHLO probe cycle aims to test the behavior of the NGFW proxy in preserving the elements and ordering a single list-type parameter highlighted in Figure \ref{chlo-fig}. The C2 server tries to decode the expected probe message from its received TLS CHLO parameter arrangement. If it decodes an expected probe message, then it implies that the existing middleboxes do not interfere with the values of the given list-type parameter. The C2 server response with different certificates to signal ACK or an NACK to the malware agent. The malware agent proceeds to the exfiltration phase only if it receives an ACK response to its probe TLS CHLO. Otherwise, it assumes that the encoding information in the probed list-type parameter is infeasible and moves on probing another parameter. If the NGFW is configured to not preserve any of the value arrangements, the malware agent should consider the possibility that the selection encoding (bitmap) outlined in Section \ref{be-a} offers decreased throughput. Discussed in Section \ref{fw-results}, it is very common for the older and ossifying middleboxes to still preserve the values of ciphers, extensions, and hashing algorithms with a different statically enforced order.}

\tclm{The malware agent can very well leverage the recent GREASE values (proposed in RFC 8701) in its probed parameter to examine the compliance of the NGFW proxy with respect to both the selection and arrangement of the parameter values.
NGFWs often employ IDS signatures, including TLS fingerprints, to detect the initialization (phone-back) packet from the disclosed malware samples to their C2 servers.
The Helol tunnel can further evade such attempts by allowing the malware agent to dynamically generate an initial arrangement of the values in the probe TLS CHLO packet.
This can be achieved by different means, including transmitting a dynamic hash digest (using the public key of the C2 server) of the probe message using the TLS CHLO parameter arrangement instead of a static probe message.}
Furthermore, there are scenarios where multiple malware agents need to reach the C2 server from the same enterprise network. To distinguish between the Helol tunnel session terminated by the same NGFW, the malware agents need to utilize an identification value embedded in their messages prior to the encoding, or in one of the TLS extensions while communicating the with same C2 server.

\subsection{Data Exfiltration}
As illustrated in Figure \ref{active-diag}, after completing the one-time NGFW probing during the initialization phase, the malware agent transmits the stolen data in chunks encoded in the arrangement of parameter values preserved through the NGFW. The size of which is calculated from the number of values used in the parameter list $\left\lfloor \log_2 N! \right\rfloor$. It can simultaneously exploit the order of elements in multiple list-type parameters within the same TLS CHLO. The malware agent uses known factoriadic methods (e.g., \cite{montanez}) to encode the binary data to the order of the parameter list while making the TLS CHLO packet. The total throughput of this method has been analyzed in Section \ref{tput} and in Appendix Section \ref{app-a}.

In a manner similar to receiving a discovery response, the C2 server can use its choice of certificate or other response attributes to ACK/NACK each or a group of consecutive TLS CHLO packets. It then proceeds in decoding the payload data chunks while the malware moves on to sending the TLS CHLO for the next chunk of data, if it receives the ACK in form of a valid server hello response from NGFW. Otherwise, it should retransmit the lost data chunk or resend a discovery probe to change its strategy. Finally, the malware agent can signal the data transfer session termination to the C2 server by including a pre-configured extension, or a fingerprint, in its last TLS CHLO packet.

\tclm{It is worth noting that sending successive TLS CHLO without proceeding past the server response does not constitute an anomalous event in most networks. In fact, it frequently happens when NGFWs perform proxy on the internet applications with anti-interception measures such as certificate / public key pinning. Such applications terminate the connection right after receiving a proxy certificate and re-attempt the TLS connection periodically.
}

\begin{table*}[ht]
\centering
\caption{Length statistics of Helol Tunnel's target TLS CHLO parameter lists}
\begin{tabular}{|p{7cm}|c|c|c|}
\hline
\textbf{NGFW Model } & \textbf{Lower Quartile (25\%)} & \textbf{Median} & \textbf{Upper Quartile (75\%)} \\ \hline
\textbf{Top Application SNI } & 8 & 11 & 15 \\ \hline
\textbf{Ciphers List (Trisul's JA3 DB \cite{trisul})} & 15 & 22 & 36 \\ \hline
\textbf{Ciphers List (local traffic capture)} & 17 & 19 & 20 \\ \hline
\textbf{Extensions List (Trisul's JA3 DB \cite{trisul})} & 5 & 6 & 10 \\ \hline
\textbf{Extensions List (local traffic capture)} & 13 & 14 & 15 \\ \hline
\textbf{Elliptic Curves List (Trisul's JA3 DB \cite{trisul})} & 2 & 3 & 11 \\ \hline
\textbf{Elliptic Curves List (local traffic capture)} & 3 & 3 & 4 \\ \hline
\textbf{Sig. Hashing Algo. List* (in local traffic capture only)} & 10 & 11 & 12 \\ \hline
\end{tabular}
\label{t1}
\end{table*}

\section{Paths to Remediation} \label{remediation-sec}

The Helol tunnel exploiting TLS cryptographic parameters to exfiltrate information has the potential to pose significant security threats to organizations with strict data loss protection policies. \tclm{We have demonstrated the trade-off between the recent anti-ossification and anti-fingerprinting measures discussed in Section \ref{ja3-sec} and the vulnerability to the Helol Tunnel covert channel exploitation. We have dedicated this document to explaining the proposed Helol tunnel and its viability and impact. Therefore, we suffice to outline the following remediation strategies to lay the foundation for future remediation research.}

\emph{Static Order of Elements:} \tclm{A straightforward approach is to make the anti-ossification measures optional, so NGFWs can still enforce a static configuration of cryptographic parameters in the TLS CHLO packet. This offers NGFW operators the tools to manage the trade-off between TLS extensibility and data protection in their networks. For example, they can configure the NGFW policies to apply static TLS CHLO configurations only in restricted network segments.}

\emph{Randomizing Order of Elements:} \tclm{In cases where enforcing a static order disrupts the critical Internet applications, the NGFWs could preserve the values in inbound TLS CHLO list-type parameters and just apply an additional layer of random permutation. This could force the Helol tunnel to lose the permutation encoding option and either fail or resort to a much slower selection encoding scheme.}

\emph{Advanced Monitoring:} \tclm{Using advanced statistical and machine-learning (ML) algorithms could be a more promising alternative to the ossifying approaches above. As mentioned in Section \ref{ja3-sec}, Helol Tunnel TLS CHLO packets could be indistinguishable from those produced by the anti-fingerprinting mechanism in the incumbent applications. However, given sufficient compute power to monitor the incumbent TLS CHLO trends, future IDS systems could potentially discover statistically significant features to identify Helol tunnel activities in such an environment.}

\section{Numerical Analysis \& Results} \label{num-sec}
In this section we formulate the throughput of the proposed Helol tunnel and the SNICat, the most comparable TLS CHLO covert channel. Then we present the relevant TLS fingerprint data sets that we can use to compute the throughput distributions and compare them.
\subsection{Throughput Analysis} \label{tput}
The expected data rate for the covert channels exploiting the TLS CHLO is calculated by multiplying these two variables: the encoded payload size per packet, and the average number of exploited TLS CHLO packets per second. We can compare the methods based on the former, while the latter depends on the victim network's condition thereby impacting all methods in the same way.

Many enterprise NGFWs do not care about TLS connections per host per second as long as they are below a certain threshold limit, which offers protection against overloading attacks. In practice, these rate threshold limits rarely flag a host sending tens (or more) of TLS CHLO per second. However, there are some emerging technologies that flag unusual network activities using statistics and ML methods \cite{acm-survey}. To avoid NGFWs using such technologies, TLS CHLO packets should be emitted at much lower rates, closer to the per-host average, to go unnoticed by NGFWs. The authors in \cite{android-tls} have collected TLS telemetry data from a large set of Android applications. Their results indicate an average of $0.48$ TLS CHLO packets per second per active device. This finding is aligned with the $0.42$ TLS CHLO packets per active user that we measured in our lab network described in Section \ref{dataset-sec}.
It is also worth noting that the passive version of the Helol tunnel presented in Section \ref{pcc-sec} is much less affected by this limit, since it can leverage encoding data in TLS CHLO packets coming from multiple network hosts.

Henceforth, we use the encoded payload size per packet as a reliable gauge to draw a throughput comparison between Helol tunnels and SNICat, the most effective TLS CHLO method presented in \cite{snicat}. Both methods encode data in the TLS CHLO packets that are intercepted and then relayed by NGFWs. Any difference in individual packet byte-size has negligible impact on the data transmission throughput mostly throttled by NGFW packet processing delays.

For a given TLS CHLO packet, the Helol tunnel method tries to encode data in the arrangement of the elements in any of these lists: cipher suites, extensions, elliptic curves, and signature hashing algorithms whose respective number of elements are denoted by $\{N_i\}_{i=1}^4$.
Then the total transferable payload size (throughput) using the factorial encoding, such as \cite{montanez}, can be computed as
\begin{align} \label{helol-tput}
T_{\text{Helol}} = \frac{1}{8}\sum_{i=1}^{4} \delta_i \left\lfloor\log_2 N_i! \right\rfloor~ \mbox{bytes/packet},
\end{align}
where $\delta_i \in \{0,1\}$ indicates if the $i$-th list of elements is exploited or not.

As indicated in Figure \ref{chlo-fig}, attackers using SNICat will be limited to using $1<N_s\leq253$ form $36$ available lower-case alphanumeric characters per TLS CHLO to encode binary data. Hence, the throughput of this method can be formulated as

\begin{align} \label{snicat-tput}
T_{\text{SNICat}} = \frac{N_s}{8}\left\lfloor\log_2 (36) \right\rfloor = 0.625\times N_s~ \mbox{bytes/packet}
\end{align}

\subsection{Evaluation Datasets}\label{dataset-sec}
In order to evaluate the throughput statistics for SNICat method and the Helol tunnel in different scenarios, we have used the datasets described below to obtain the distributions of the number of elements ($N$ values) used in \eqref{snicat-tput} and \eqref{helol-tput}.

\emph{Top Application SNI:} we have evaluated $3907$ SNI strings from the traffic captured from automated crawling of the top 50 SimilarWeb websites \cite{crawler}
and the traffic captured from the top mobile applications mentioned in 2023 report by Sandvine Inc \cite{sandvine}. The first row in Table \ref{t1} provides the key quartile statistics for the distribution of $N_s$.

\emph{Public JA3 Fingerprints:} There are several public databases of JA3 TLS fingerprints from both application and malware, such as the \texttt{ja3.me} and \texttt{ja3print} tool by Trisul Network Analytic\cite{trisul}. Table \ref{t1} also shows the quartile statistics on the number of cipher suites, extensions, and elliptic curves of 364 common application fingerprints in Trisul's JA3 database \cite{trisul}, excluding those from malware and security research tools. JA3 records can only help estimate throughput in \eqref{helol-tput} without exploiting signature hashing algorithms, since that is not being fingerprinted.

\begin{figure}[htb] \centering
    \includegraphics[width=7.5cm]{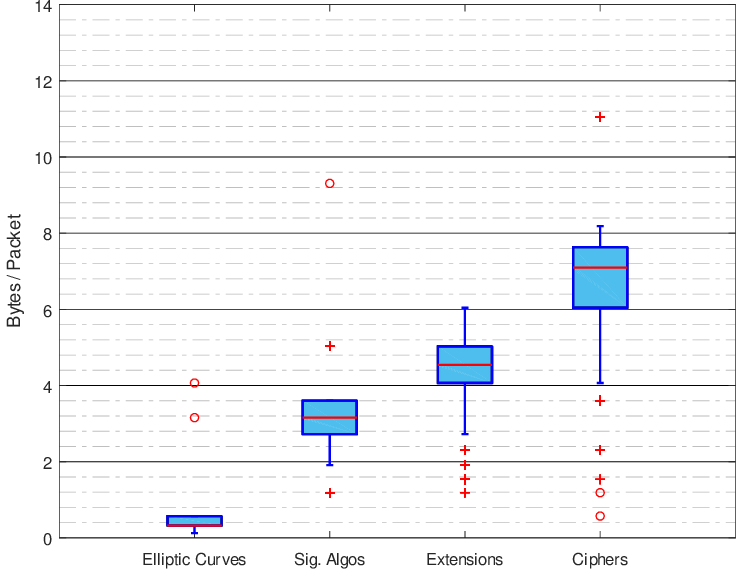}
		\caption{Permutational encoding throughput using different TLS CHLO parameter types, based on the length distributions observed in the local traffic capture.}
    \label{passive-tput-fig}
\end{figure}

\emph{Local Traffic Captures:} To circumvent the gaps in JA3 fingerprints, we have captured TLS CHLO packets from all devices in a Keysight Technologies lab for $4$ consecutive days with $2$ active enterprise personnel during $8$ business hours. Devices include: $2$ Microsoft Windows laptops, $2$ Google Android mobile phones, $2$ Apple iOS devices, $3$ IoT cameras and a Linux desktop. There were $96825$ TLS CHLO packets, where only $137$ of them showed unique combinations of cipher suites, extensions, elliptic curves and hashing algorithms without accounting for their permutations. As observed in \cite{android-tls} data, the TLS CHLO generation rate varies widely among the applications. The web-browsers and enterprise applications had more recurrence in our dataset. Also included in Table \ref{t1} statistics, the values for the local TLS CHLO captures show less variance (closer quartiles) compared to their counterparts in \cite{trisul}, where each application fingerprint is counted only once.
Using the local traffic capture data, the boxplot in Figure \ref{passive-tput-fig} demonstrates the throughput distribution using each TLS CHLO parameter. These distributions are formed by plugging the length for each of the observed TLS CHLO parameter lists into \eqref{helol-tput}. It can be seen that the number of cipher suites is the most contributing factor in the throughput followed by the extensions, hashing algorithms and a minuscule share from Elliptic Curves.
Therefore, it is important for malware agents to prioritize the exploitation of the parameters in the same order of throughput gains.
\begin{figure}[htb] \centering
    \includegraphics[width=7.5cm]{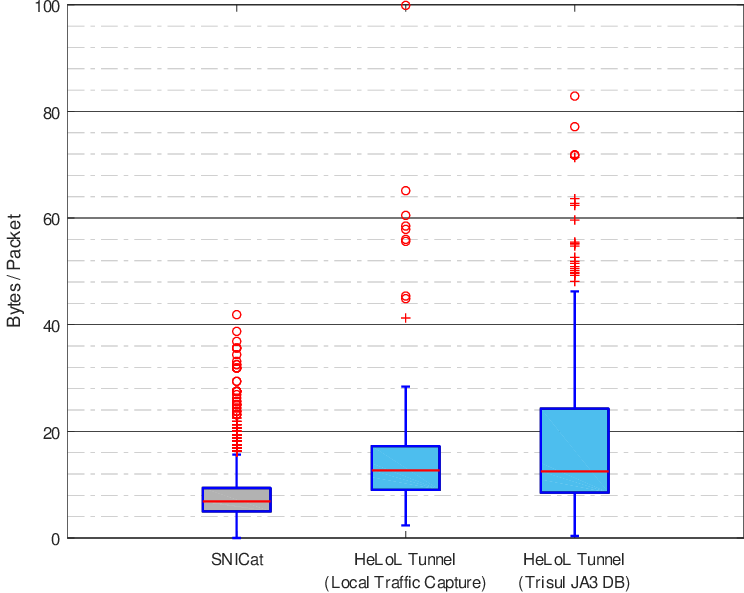}
		\caption{Helol tunnel throughput v.s. SNICat scheme. }
    \label{tput-comp-fig}
\end{figure}
\subsection{Comparison Results}

A malware agent exploiting the Helol tunnel can populate its lists of targeted parameters similarly to any trust application fingerprints from datasets such as Trisul's in \cite{trisul}. As long as the application is not among those blocked by the NGFW, sending TLS CHLO with permutation encoded parameters at around dozens per second should generally not raise any intrusion warnings. Plotted in Figure \ref{tput-comp-fig}, we have derived the throughput distribution for the channel, matching trusted application fingerprints from Trisul's dataset. There we have also included the Helol tunnel throughput distribution using any of the $137$ distinct TLS CHLO configuration observed in our local traffic capture. There we can see the proximity of the throughput median computed from either dataset.

Finally, in gray, is the throughput distribution for the SNICat method in Figure \ref{tput-comp-fig} using the length values in the dataset of the top applications inserted in \eqref{snicat-tput}. We can observe significant throughput gains from Helol tunnel scenarios compared to SNICat in terms of median, third, and fourth quartiles. Furthermore, comparing the SNICat quartiles with the Helol tunnel throughput quartiles per parameter in Figure \ref{passive-tput-fig}, it can be seen that the Helol tunnel retains its throughput advantage over this method even when exploiting cipher list or extensions for permutational encoding.

\section{Conclusion} \label{conc-sec}
In conclusion, this research has demonstrated the potential to exploit the combinatorial properties of cryptographic parameters in the TLS Client Hello for data exfiltration. The novel Helol tunnel method proves to be an effective covert channel technique that exploits the very first packet in the TLS protocol handshake, leading to successful evasion of modern Next-Generation Firewalls with advanced TLS proxy and data protection. Furthermore, Helol tunnel is applicable in versatile attack scenarios including the passive scenario, where hosts can communicate without acquiring an IP address from the incumbent network.
This highlights the level of risk posed by this method compared to the other state-of-the-art covert channels.

Looking ahead, the proposed exfiltration technique opens new avenues for further research. The mitigation strategies suggested in this study involve NGFW reconfiguration of TLS CHLO packets that can degrade the performance of network applications, resulting in a trade-off between TLS extensibility and the level of data protection. This can be a new research opportunity to explore alternative methods of disrupting Helol tunnel without enforcing strict cryptographic TLS configuration using emerging ML models in the field of TLS fingerprinting.

\appendices

\section{Efficacy of Combinatorial Encoding Schemes}\label{app-a}

Given a set of $N$ symbols $\mathcal{S}\triangleq\{s_1,s_2,s_3,\cdots, s_N\}$, Alice communicates with Bob using codewords (messages) from a predefined \textit{codebook} set $\mathcal{C}\triangleq\{\bf{c}_1, \bf{c}_2,\cdots, \bf{c}_M\}$ of size $M$. Each codeword is a sequence of symbols from $\mathcal{S}$. In case of exploiting list-encoded protocol parameters, the codewords cannot contain repeating symbols, thus their length will remain less than $N$.

Bob decodes each received codeword (assuming error-free decoding) to recover the binary data transmitted by Alice. The channel capacity (maximum throughput) is defined as the maximum number of data bits representable by each codeword. This quantity is $\log_2 M$ for the given codebook of size $M$. However, each symbol in the codeword also has to be transmitted in binary format within the underlying protocol packets that requires at least $\log_2 N$ of bits per symbol. In addition to the complexity of the decoding operation, often $O(N)$, the \textit{efficiency factor} of a codebook , $\eta$, can be defined as the ratio of its capacity to the expected (average) bits transmitted per codeword. \tclo{This quantity is analogous to the steganographic capacity used in \cite{heinz-tls-1}.}
\begin{align}\label{etadef}
\eta \triangleq \frac{\log_2 |\mathcal{C}| }{E(|\textsc{c}|) \times \log_2 N}~,
\end{align}
where $E(|\textsc{c}|)$ denotes the expected number of symbols in each codeword $\textsc{c} \in \mathcal{C}$. The above efficiency factor is a unitless quantity between $0$ and $1$ analogous to the concept of coding rate in error-correcting codes. The higher values of $\eta$ indicate faster data transmission and better bandwidth utilization. It is straightforward to show that maximum value $\eta_{max} = 1$ is achieved when the codebook includes all possible symbol sequences of length $N$ without the constraint on repeating symbols, i.e.
\begin{align}
C_{max} = \{\forall~\textsc{c} = [c_1,c_2,\cdots,c_N]~|~c_i \in \mathcal{S}\},~~|\mathcal{C}_{max}| = N^N \nonumber
\end{align}
In the following we derive $\eta$ for possible encoding schemes relevant to the Helol tunnel scenarios and demonstrate the merit of our choice of permutational encoding.

\subsection {Selection (Bitmap) Encoding} \label{be-a}
If the order of the symbols in Alice's codewords are intercepted and sorted by a Warden (NGFW) en route to Bob, their order information is lost. Thus, any pair of codewords in $\mathcal{C}$ should differ by at least one symbol or more. The power set (the set of all subsets) of $\mathcal{S}$ can be used as the largest possible codebook in this scenario. They can employ the following \textit{bitmap encoding} scheme:
\begin{itemize}
\item Alice encodes blocks of $N$ data bits $[b_1,b_2,\cdots,b_N]$ to a codeword at a time, and scans every element of the data block and includes $s_i$ in the codeword only if $b_i = 1$.
\item Upon receiving a codeword, Bob initializes every bit in its receiver buffer $[b'_1,b'_2,\cdots,b'_N]$ to zero. Then it scans over each symbol in the codeword and sets $b'_i = 1$ only if it encounters $s_i$.
\end{itemize}
Given the equal probability of the input binary data streams, one can show that $E(|\textsc{c}|) = N/2$ that yields the efficiency of this scheme to be
\begin{align} \label{etabe}
\eta_{BE} = \frac{2}{\log_2 N}
\end{align}

\subsection {Permutation Encoding } \label{pesub}
Using the classic algorithms explained in \cite{montanez}, Alice encodes blocks of $\left\lfloor \log_2 N! \right\rfloor$ data bits into a permutation of the entire symbols in $\mathcal{S}$ to produce a codeword of size $ N$, which Bob later decoded to recover the same block of bits. In this case, the codebook consists of $N!$ codewords of equal sizes, so the efficiency of this scheme calculated as
\begin{align}\label{etape}
\eta_{PE} = \frac{\log_2 N! }{N\log_2 N } \approx 1 - \frac{\log_2 e}{\log_2 N} + O(\frac{1}{N}),
\end{align}
where applying the Stirling's formula for $N!$ with large $N$ reveals its steady convergence to the maximum $\eta_{max} = 1$

\tclg{
\subsection{Partial Permutation Encoding}
Given the non-repeating symbols constraint on the codewords, one may compile a codebook of all partial permutation sequences from $\mathcal{S}$. Regardless of its more sophisticated encoding / decoding algorithms, we can derive its efficiency for comparison to \eqref{etape}. Since each codeword is a permutation of a subset of $\mathcal{S}$, the codebook can be expressed as
\begin{align}
\mathcal{C}_{PPE}=\{~\textsc{c}_{km} = perm_k(S_m)~|~S_m \subseteq \mathcal{S} ,~1 \leq k \leq |S_m|!~\}.\nonumber
\end{align}
Therefore, we can calculate the codebook size as follows:
\begin{align}
|\mathcal{C}_{PPE}| &= \sum_{m=1}^{N} \binom{N}{m} m! =  N! \sum_{j=1}^{N-1}\frac{1}{j!} \approx \left\lfloor eN! \right\rfloor
\end{align}
Since the codewords are considered equiprobable, the expected length can be derived using the above approximation techniques
\begin{align}
E(|\textsc{c}|) &= \sum_{m=1}^{N} m \frac{\binom{N}{m} m!}{\left\lfloor eN! \right\rfloor} = \frac{N!}{\left\lfloor eN! \right\rfloor}\sum_{m=1}^{N}\frac{m}{(N-m)!} \nonumber\\
                &= \frac{1}{e}\sum_{k=0}^{N-1} \frac{N-k}{k!} \approx N-1
\end{align}
Now we have all of the components to evaluate the efficiency factor from \eqref{etadef} and discover its relationship with $\eta_{PE}$ in \eqref{etape} as follows:
\begin{align}\label{etappe}
\eta_{PPE} = \frac{\log_2 e + \log_2 N!}{(N-1)\log_2 N} \approx (1 + \frac{1}{N})\eta_{PE}
\end{align}
}

\tclm{
\subsection{Advantage of Permutation Encoding}
Reviewing the calculated coding efficiencies, one can see that the efficiency of the permutation encoding in \eqref{etape} approaches the maximum by increasing the codeword length $N$. On the other hand, the bitmap efficiency factor in \eqref{etabe} does not scale properly with respect to the codeword length and rapidly declines. The partial permutation encoding uses a more sophisticated codebook with variable-length codewords requiring much more complex decoding scheme than the proposed fixed-size permutation encoding. However, it scores negligible throughput gains of $\left\lfloor\log_2 e\right\rfloor$ over the permutation encoding, which translates to an extra bit per codeword. In addition, its efficiency in \eqref{etappe} rapidly converges to the efficiency of permutation encoding. Therefore, we conclude that the proposed fixed-length permutation encoding is the optimal choice for the Helol tunnel given the encoding constraints mentioned above.
}

\begin{table*}[ht]
\centering
\caption{\tclg{Empirical evaluation of some commonly deployed enterprise NGFWs against Helol tunnel}}
\begin{tabular}{|p{7cm}|c|c|c|}
\hline
\textbf{NGFW Model } & \textbf{Model A} & \textbf{Model B} & \textbf{Model C} \\ \hline
\textbf{ Release Year of Tested NGFW Software} & 2023 & 2020 & 2024 \\ \hline
\textbf{Permutation Encoding of Ciphers} & Exploitable & No & Exploitable \\ \hline
\textbf{Permutation Encoding of Hash Algo} & Exploitable & No & Exploitable \\ \hline
\textbf{Permutation Encoding of EC Curves} & Exploitable & No & Exploitable \\ \hline
\textbf{Permutation Encoding of Extensions} & No & No & Exploitable \\ \hline
\textbf{Bitmap Encoding w/ selection of TLS params } & Exploitable & Exploitable & Exploitable \\ \hline
\end{tabular}
\label{t2}
\end{table*}

\tclg{
\section{Evaluating Next-Generation Firewalls} \label{fw-results}
We have implemented a proof of concept of the common Helol tunnel scenario at our Keysight BreakingPoint lab to evaluate 3 commonly deployed enterprise NGFW models, whose evaluation results are anonymously compiled in Table \ref{t2}. We have enabled maximum security features in each product, including advanced TLS proxy and data leak prevention. The results show that 2 out of 3 have complied with anti-ossification measures that leave them vulnerable to the Helol Tunnel permutation encoding. All 3 of these products retain the values in their inbound TLS CHLO packets as long as they are within the set of values recommended by NIST. This will make all 3 vulnerable to the low-throughput bitmap encoding outlined in Section \ref{be-a}. We also did not observe any of the products raising any warning about our simulated malware agent sending repetitive TLS CHLO packets.
}

\section{Helol Tunnel for Passive Covert Channels} \label{pcc-sec}

Presented in \cite{passive-2004} two decades ago, passive covert channels took a very compelling approach to hiding the identity of communicating peers. However, most of these early works focused on exploiting layer 4 features, which are no longer relevant in the age of modern NGFWs. \tclo{The authors in \cite{heinz-tls-1, heinz-tls-2} have considered this scenario in their proposed TLS covert channels, most of which are not compatible with interactive TLS proxies}. Our preliminary studies show that the Helol tunnel is likely the only method capable of establishing such TLS CHLO covert channels across NGFW with interactive proxies.
\begin{figure}[ht] \centering
    \includegraphics[width=7.5cm]{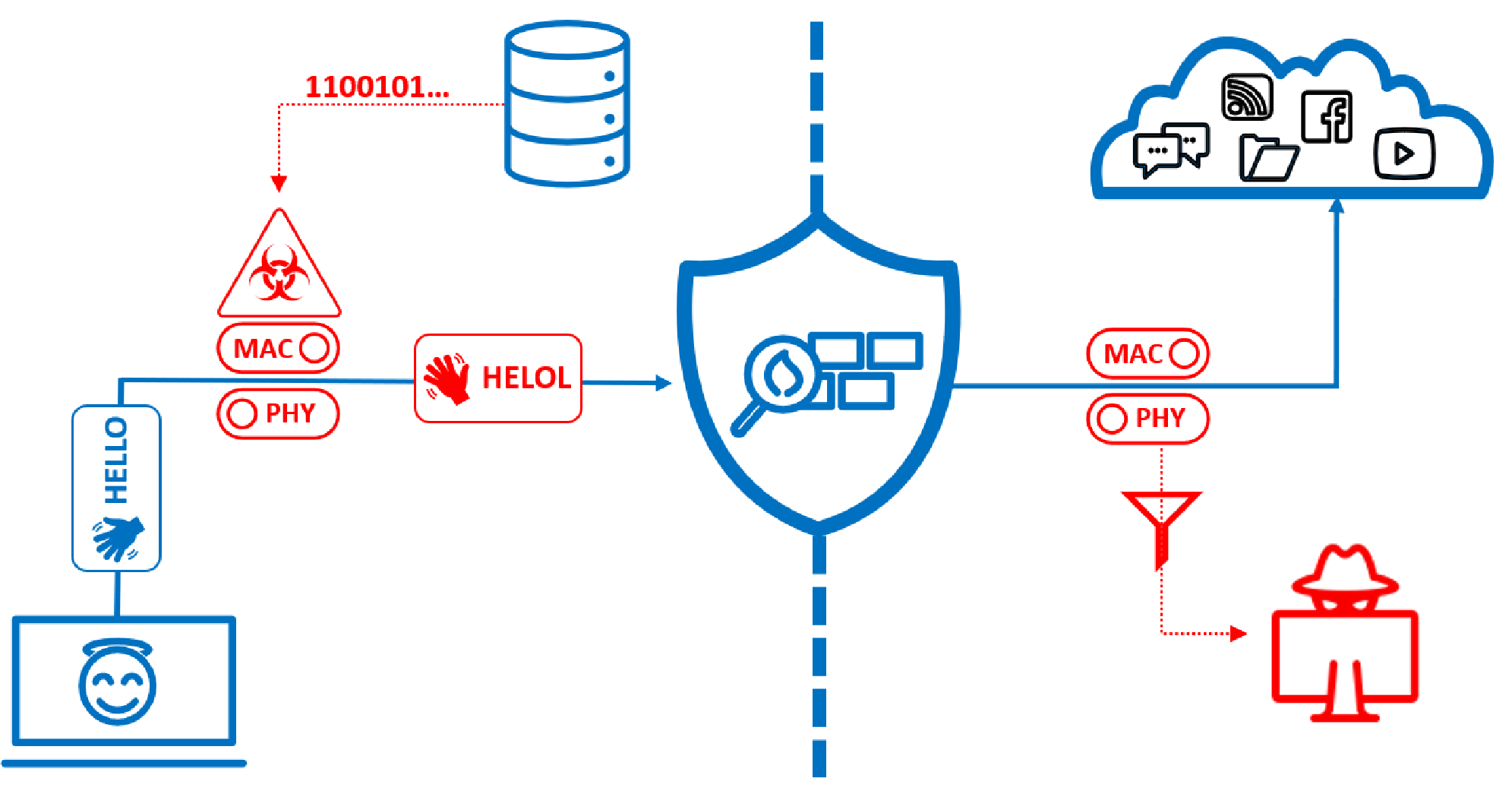}
		\caption{Proposed passive covert channel exploitation}
    \label{pcc-fig}
\end{figure}

This scenario is different from the common channel scenario used by the malware agents in Figure \ref{active-diag}. It involves a covert channel between two compromised network equipment devices with tap access to the layer 3 route from the victim network hosts to the internet. Referring to the diagram in Figure \ref{pcc-fig}, the insider agent can be an infected LAN switch, or malicious software with privileged access to the network interface of a victim host that is connected to the Internet through the NGFW. The outsider agent can passively monitor (wiretap) NGFW outbound traffic to the Internet. It can be another infected LAN switch installed in the backhaul service provider’s infrastructure. It is worth noting that neither agent needs an IP address as they can only capture incumbent TLS packets and modify them en route to their destinations.
The insider agent wishes to communicate covertly with its outside counterpart without revealing its presence in the victim’s LAN. Hoping to be observed and decoded by the outsider agent, it aims to encode its message in the TLS packets that flow out of the network without disrupting the incumbent TLS connections. This helps both agents remain hidden from any layer 3 to 7 inspection for long periods of time. Hence, this type of covert channel is more prevalent in the realm of supply-chain attacks.
Thus, Our proposed method of encoding information in the ordering of the TLS CHLO parameters can be a viable solution to achieve this despite the proxy in layer 4 by non-ossifying TLS proxies.



\tclb{
\section{Responsible Disclosure \& Ethics}\label{ethics}
In compliance with the high ethical standards of security research, we have responsibly disclosed the Helol Tunneling covert channel scheme to the impacted network security vendors prior to any publication attempts. We have shared our research materials in the Coordinated Vulnerability Disclosure (CVD) program offered by the US Cybersecurity and Infrastructure Security Agency (CISA), where we privately discussed its viability and impact with the vendors participating in the program. As of November 2024, they have lifted the publication embargo for this research. We remain committed to working with vendors and standardization entities who are interested in developing countermeasures against this covert channel scheme.
}
\bibliographystyle{IEEEtran}
\bibliography{Helol-sparrow}
\end{document}